*HE.4.2.17*

# The South Atlantic Magnetic Field Anomaly and Its Effect on the Calculated Production of Atmospheric Neutrinos

J. Poirier

*Physics Department, 225 NSH, University of Notre Dame, Notre Dame, IN 46556, USA*

## Abstract

The theoretical calculations of the production of neutrinos via cosmic rays incident upon the earth's atmosphere (Barr, Gaisser, & Stanev, 1989; Becker-Szendy et al., 1992; Bugaev & Naumov, 1989; Gaisser, Stanev, & Barr, 1988; and Honda et al., 1995) are examined. These calculations use a one-dimensional approximation in the production, transport, and decay of the produced particles. Examined are various additional effects of the earth's magnetic field and the three-dimensional nature of the problem which have the effect of decreasing the calculated ratio of muon neutrinos to electron neutrinos. This would shrink the disparity between theory and the Super-Kamiokande experimental results (Y. Fukuda et al., 1998) and make the neutrino oscillation hypothesis less compelling.

The earth's magnetic dipole field repels primary charged cosmic ray particles (either positive or negative) incident upon this field from its exterior. If their energy (more precisely, their momentum divided by their charge, $p/Z$) is below a cutoff energy (which depends on geomagnetic latitude), then these primaries cannot penetrate the magnetic field sufficiently to reach the surface of the earth. The cutoff energy depends upon the incident direction, the geomagnetic latitude, and the primary's impact parameter, as well as $p/Z$. Charged primaries with energies below this cutoff energy (or momentum) are repelled such that, at their distance-of-closest approach (H_min), they are above the surface of the earth. H_min increases as the primary energy decreases. At H_min, the primaries are moving tangent to the earth's surface. Since the particles are nearly tangent to the surface of the earth, they can travel large distances through air of low density and still have an appreciable chance to interact if H_min is not too large. The scale length for a hadron to interact in air is 90 g/sq-cm. Upon interacting, the primaries produce mainly pions at small angles; thus the pions are also nearly tangent to the earth's surface. The pions decay to a muon and $\nu_\mu$, both at small angles relative to the pion direction. The pions decay in a short time; their lifetime at rest, $\tau_0$, is .026 microsec, or $c\tau_0$ is 7.8 m. At a typical energy of 4 GeV, their lifetime is .75 microsec and their mean travel distance is 220 m. The $\nu_\mu$ is also nearly tangent to the curved surface of the earth at this point of minimum proton height and thus is incapable of being detected by earth-bound neutrino detectors.

Figure 1 depicts a proton primary of relatively low energy being repelled by the earth's magnetic field. At the point of nearest approach, the proton, p, is tangent to the earth's surface and is skimming the top of the earth's blanket of air (A, like nitrogen) with which it can interact. The interaction produces mainly multiple pions which, as noted above, quickly decay to a muon and $\nu_\mu$.

$$p\,A \rightarrow N^+\pi^+ + N^-\pi^- + N^0\pi^0 + A \qquad (1)$$
$$\pi^+ \rightarrow \mu^+ + \nu_\mu \qquad (2)$$

where $N^+$, $N^-$, and $N^0$ are respectively the numbers of $\pi^+$, $\pi^-$, and $\pi^0$ pions produced, and A is an atomic nucleus of air, like nitrogen. As seen in the figure, the $\nu_\mu$ (2) from the pion decay is directed away from any earthbound detectors.

On the other hand, the decay muon lives longer than the pion; the muon lifetime at rest, $\tau_0$ is 2.2 microsec, or $c\tau_0$ is 660 m. Upon pion decay, the muon takes the major fraction of the momentum of the parent pion; to compare with the pion, a typical energy of 3.5 GeV is used. At this energy, the muon's lifetime is 73 microsec and its mean travel distance is 22 km. If before its decay the muon scatters in the atmosphere or bends in the earth's magnetic field (30 times more probable than the short-lived pion), then its decay neutrinos are accessible to earthbound detectors. The muons have less momentum than the proton primary parent (which is nearly trapped in the earth's magnetic field), so the muon *is* trapped in this magnetic field which increases the probability of the muon to bend in the field so as to point toward earthbound detectors. In addition, the $\bar{\nu}_\mu$ and $\nu_e$ from the muon decay

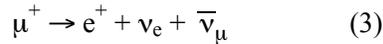

$$\mu^+ \rightarrow e^+ + \nu_e + \bar{\nu}_\mu \qquad (3)$$

have decay angles that can add to the muon's magnetic bend or scattered angle.
Figure 1 shows that if the $\mu^+$ angle is bent and/or scatters to a large enough angle, then the $\bar{\nu}_\mu$ and $\nu_e$ from (3) can point toward earthbound detectors.

The primary cosmic rays considered here have not been included in the Monte Carlo (MC) calculations as these primaries are those that do not reach the surface of the earth. The pion decay $\nu_\mu$ is lost and the $\bar{\nu}_\mu$ and $\nu_e$ from the muon decay do have a chance of pointing toward an earthbound detector. The $\nu_\mu$ to $\nu_e$ ratio from this added source of cosmic rays is thus reduced from the customary ratio of 2:1 to 1:1 since one of the $\nu_\mu$-s in the $\pi^- \rightarrow \mu^- \rightarrow e$ decay chain cannot be detected.

This general effect occurs everywhere above the surface of the earth. However, the effect is maximum at the South Atlantic Anomaly (SAA), a region where the earth's magnetic field has a minimum of intensity. At this lower magnetic field value, the low energy cosmic primaries can penetrate nearer to the surface of the earth; that is, H_min is lower at the SAA than any other place on the earth's surface (at the same or smaller geomagnetic latitude). The cosmic ray flux of hadrons is very peaked toward low energy with a differential flux, dF, given by:

$$dF = c * E^{-2.7} dE \qquad (4)$$

This region of lower magnetic field, B, allows lower energies to reach the earth's atmosphere; since the intensity of lower energies is much higher, then the cosmic ray intensity is also much higher at the SAA. For example, compare the SAA region's flux of cosmic rays to another region of the earth like Japan where B is two times higher. The resulting flux, dF/dE, would then be six times higher at the SAA compared to Japan. This SAA region of maximum cosmic ray intensity (because of the minimum magnetic field) is near Brazil and Argentina in South America. The SAA is almost directly on the opposite side of the earth from Japan, the location of the Super-Kamiokande experiment. This flux of neutrinos comes in addition to the ones in the standard MC calculations. This added flux has an excess of $\nu_e$ and thus would give a high value of $\nu_e$ (or a minimum in the $\nu_\mu$ to $\nu_e$ ratio) coming from the SAA--these would be the upward-going neutrinos for the Super-Kamiokande experiment on the opposite side of the earth. Super-Kamiokande sees 1) a minimum for R for the upward-going neutrinos relative to the downward-going neutrinos, and 2) an excess of $\nu_e$. Super-Kamiokande's downward-going neutrinos would also be affected, but less so since the earth's magnetic field is stronger there and the low energy cosmic ray flux correspondingly lower.

The above discussion argues for an additional source of neutrinos with a depleted ratio of $\nu_\mu/\nu_e$ by a factor of two due to the SAA. Another additional source of cosmic ray secondary neutrinos is the Van

Allen Radiation Belt, VARB, a region of protons and electrons stored in the magnetic field of the earth. Consider some reactions of these protons and electrons:

$$e^- + p \rightarrow n + \nu_e \quad (5)$$
$$p + A \rightarrow N^+\pi^+ + N^-\pi^- + N^0\pi^0 + A \quad (6)$$

where "p" in (5) are protons in the air nuclei, "A" in (6) is an air nucleus like nitrogen, and $N^+$, $N^-$, and $N^0$ are the numbers of $\pi^+$, $\pi^-$, and $\pi^0$ that are produced in the interaction. Reaction (5) gives an excess of $\nu_e$-s with *no* accompanying $\nu_\mu$-s. In addition, the $\nu_e$ has about three times its antiparticle's cross section for being detected in the neutrino detectors. Reaction (5) is a weak interaction, so the cross section is small; however, the intensities of electrons in this stored belt are extremely high compared to cosmic ray rates. It is in the region of the SAA that these stored particles come closest to the region of denser air where the chance of an interaction is the highest. Magnetic disturbances from the sun can dump large numbers of e's and p's from the VARB to the atmosphere in this SAA region. Reaction (5) is thus an additional source of $\nu_e$-s at the SAA (with no accompanying $\nu_\mu$-s) which would thus depress the ratio of $\nu_\mu$ to $\nu_e$ overall.

Reaction (6) is a proton primary like a cosmic ray primary, so it would, on the surface, give the expected ratio of $\nu_\mu$ to $\nu_e$. However, it is argued that for energies in the region of 2 to 5 GeV, $N^+$ is $\geq 2$ times $N^-$ (see the experimental references in Poirier, 1999). Since 1) the $\pi^+ \rightarrow \mu^+ \rightarrow e^+$ decay chain produces electron-neutrinos only of the particle-type whereas 2) the $\pi^- \rightarrow \mu^- \rightarrow e^-$ decay chain produces only the antiparticle-type, and 3) the cross section for detection of the $\nu_e$ is three times that of its antiparticle in this energy range, then the average detected ratio of $\nu_\mu$ to $\nu_e$ is reduced due to the excess of $\pi^+$ over $\pi^-$ from this Van Allen Radiation Belt source of protons. This argument applies to all positively charged cosmic ray primaries regardless of their origin; it is discussed further in reference (Poirier, 1998).

The actual numerical calculation of the combination of the pion's production angle, the muon's decay angle, the muon's scattering in the atmosphere, deflection in the earth's magnetic field, and the decay angles of the neutrinos relative to the parent muon is complicated. A correct calculation would require a Monte Carlo (MC) program with all of these physics details incorporated, including the three-dimensional nature of the earth. These details would be in addition to the considerable work and long times that it has already taken to incorporate the standard effects of hadronic cosmic rays in the earth's atmosphere. It is a complicated problem beyond our present capability to implement. This paper merely indicates several details which may be important in the MC programs and notes their absence from these calculations at the present time. Thus it seems premature to accept the conclusion of "neutrino oscillations" based on Super-Kamiokande's experimental results being lower than the corresponding MC calculations which are deficient in details that would lower their theoretical result. The inclusion of the effects mentioned here as well as other details (see, for example, Poirier, 1999) could bring theory and experiment together without invoking the neutrino oscillation mechanism.

Thanks to J. LoSecco for useful discussions, A. Garcia for helpful insights, and W. J. Carpenter, T. F. Lin, and A. Roesch for technical assistance.

Project GRAND is presently being funded by grants from the University of Notre Dame and private donations.

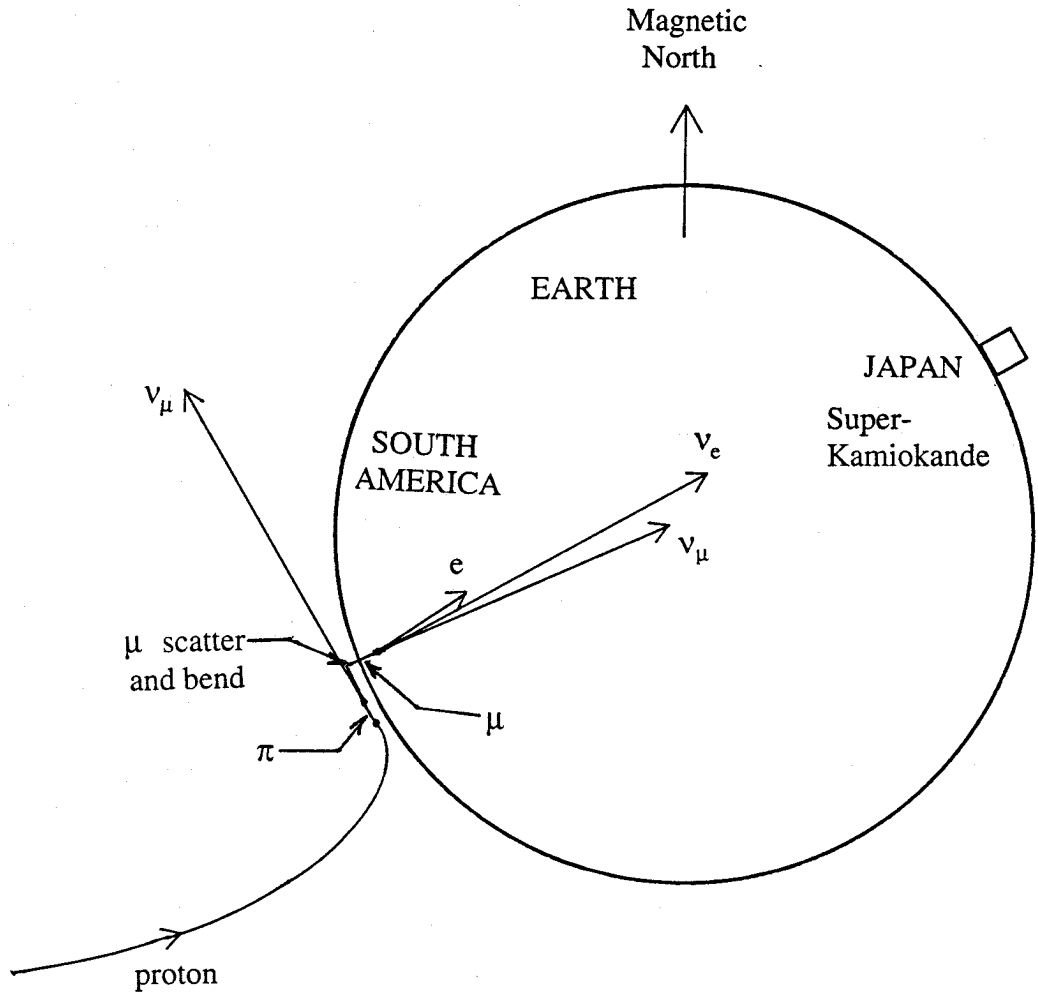

**Figure 1:** A low energy cosmic ray primary proton as it is repelled in the earth's magnetic field. At minimum height above the earth's surface the proton is parallel to the earth's surface. The proton interacts with the atmosphere creating pions which decay to muons and $\nu_\mu$'s; these $\nu_\mu$'s cannot be detected by earthbound detectors. The muons are at a lower momentum and are trapped in the earth's magnetic field where they bend and scatter. The muons then decay to e, $\nu_e$, and $\nu_\mu$. These latter neutrinos can be directed toward earthbound detectors giving a theoretical detected ratio $\nu_\mu:\nu_e$ reduced from 2:1 to 1:1.